\documentclass[doublespacing]{revtex4-1}

\usepackage{amsthm,amsmath}
\usepackage{graphicx}

\begin{document}

\title{Electronic and optical properties of Mn impurities in ultra-thin ZnO nanowires: insights from density-functional theory}

\author{Andreia Luisa da Rosa}
\affiliation{Instituto de F\'{\i}sica, Universidade Federal de Goi\'{a}s,Campus Samambaia, 74690-900 Goi\^{a}nia, Goi\'{a}s, Brazil}
\affiliation{Bremen Center for Computational Materials Science, University of Bremen, Am Fallturm 1, 28195 Bremen, Germany}
\author{Leticia Lira Tacca}
\affiliation{Instituto de F\'{\i}sica, Universidade Federal de Goi\'{a}s,Campus Samambaia, 74690-900 Goi\^{a}nia, Goi\'{a}s, Brazil}     

\author{Erika Nascimento Lima}
\affiliation{Universidade Federal de Mato Grosso, Campus Rondon\'opolis, 78005-050 Rondon\'opolis, Mato Grosso, Brazil}
\author{Thomas Frauenheim}
\affiliation{Bremen Center for Computational Materials Science, University of Bremen, Am Fallturm 1, 28195 Bremen, Germany}

\begin{abstract}
  In this work we have employed density-functional theory with
  hybrid functionals to investigate the atomic and electronic
  structure of bare and hydrogenated Mn doped ZnO nanowires with small
  diameter.  We determine changes in magnetic and electronic structure
  of Mn-doped ZnO nanowires due to surface effects, such as hydrogen
  adsorption on the surface, presence of oxygen vacancies and dangling
  bonds.  In the absence of passivation on the nanowire surface, the
  manganese atoms segregate to the surface, whereas under hydrogen
  adsorption the incorporation of Mn is energetically more favourable
  at inner sites. The presence of additional oxygen vacancies does not
  produce signficant changes in magnetic moments, although it produce significant changes in charge localization.

\end{abstract}

\maketitle

\newpage
\section*{Introduction}

Controlled incorporation of magnetic impurities in a semiconductor
provides a means to manipulate magnetic and electronic interactions.
Among transition metal ions, Mn has received long-standing interest as
a dopant in semiconductors because of its large spin magnetic moment
and therefore applications in spintronics\cite{Dietl}.  ZnO (zinc
oxide) is a wide band gap semiconductor which can be sinthesized in
several nanostructured forms. Indeed ferromagnetism due to Mn
impurities above room temperature has been reported by several
authors\,\cite{Rao2003,Theodoropoulou2006,Norberg2004,Blythe2004,Gamelin2006,Sluiter2005}.
although it has also been assigned to the formation of inclusions or
secondary phases\,\cite{Garcia2005}, while other
investigations suggested a paramagnetic
behavior\,\cite{Rao2005,Han2003,Lawes2004,Kolesnik2004}.

Mn incorporation in ZnO nanostructures has also drawn attention
\,\cite{Sapkota2016,Nanomaterials2017,ChangNL,Willander2017}.  The pioneer work by
Nair {\it et al.}\cite{Nair2006} reported ferromagnetism in diluted
Mn-doped ZnO nanowires at temperatures up to 400\,K and atributed it
to an interplay between Mn doping and native point defects. More
recently it has been assigned to the formation of magnetic polarons
\,\cite{Sapkota2016}. Nnanowire based field-effect transistor
demonstrated the presence of ferromagnetism above room temperature,
suggesting that quantum confinements improves the Curie
temperature\,\cite{ChangNL}.  Theoretical calculation suggested that
Mn incorporation can be achieved, but the interaction between Mn atoms
leads to an anti-ferromagnetic or paramagnetic
behavior\,\cite{Nanomaterials2017,Shi01,daRosa}. More recently, the
interest in Mn-doped ZnO has risen due to possible applications
in quantum information\,\cite{Moro,George}. It has been shown that
inserting Mn in ZnO nanocrystals with small diametres with passivated
surfaces leads a very efficient source of spin
decoherence\,\cite{EPL}.

Despite all the efforts, some aspects of Mn incorporation in ZnO
nanostructures remain unclear. Due to the large surface/volume ratio,
effects of surface passivation by impurities or ligands can be crucial
for successful doping, as discussed in
Refs.\,\cite{Erwin2005,Dalpian2013}. In this work, we address Mn
insertion in ultra-thin ZnO nanowires using density functional theory
\cite{Hohenberg} and hybrid-density functionals\cite{HSE}. Such wires
serve as ideal systems to investigate quantum confinement, since their
diameter is approximately 1\,nm, less than the ZnO bulk exciton Bohr
radius of 2.34\,nm\cite{Gu}.  We determine changes in magnetic and
electronic structure of Mn-doped ZnO nanowires due to surface effects,
such as hydrogen adsorption on the surface, presence of oxygen
vacancies and dangling bonds.

We show that the impurity prefers to sit at bulk positions when the
surface is adsorbed by hydrogen. On the other hand, bare wires suffer
from self-purification problems leading to segregation of the dopant
of Mn towards surface sites. Furthermore we find that Mn atoms in ZnO
create additional states in the nanowire band gap. However, for bare
wires, these states strongly hybridize with the surface states. On the
other hand hydrogenated wires passivate surface dangling bonds, which
appear to be beneficial for Mn incorporation. Finally we suggest that
oxygen vacancies do not introduce significant changes in the overall
magnetic properties. However, depending on the vacancy location, the
overlap between Mn states and surface states are affected by such
defects.

\section*{Methodology}

Density-functional theory (DFT)\,\cite{Kohn:65} and the projected
augmented wave (PAW) method, as implemented in the Vienna Ab initio
Simulation Package (VASP)\,\cite{Kresse:99}, has been used to
investigate Mn incorporation in ultra-thin ZnO nanowires. In order to
model Mn impurities in ZnO we built up an isolated, infinite wire
along the [0001] drirection containing 48 ZnO-units and a vacuum region
of 15\,{\AA}. This gives a Mn concentration of 2.1\%. To ensure
convergence of structural, electronic and magnetic properties, a
cutoff of 400 eV was used for the wave function expansion in plane
waves. Atomic forces on all atoms were converged up to
0.001\,eV/{\AA}. Brillouin zone integration of the charge density has
been done with a $(1\times 1 \times 4)$ {\bf k}-point sampling.  The
relaxation of the nanowires was done using the PBE\,\cite{Perdew:96}
functional.  
One of the key aspects in the identification of dopant states is the
proper description of the band gap of the host material. While the PBE
functional is very reliable to describe atomic properties, the same
has limitations to describe electronic structure and optical
properties in semicondutors\,\cite{Klimes_Kresse:14,Onida}. Recently the use
of modern exchange-correlation functionals in density-functional
theory has shown to yield band gap values close to
experiment\,\cite{Paier:08,Lany:10,Clark:10}. We have recently shown that the use of
hybrid functionals can reproduce the ZnO band gap and provide  correct
location of energy levels of 3$d$ and 4$f$ in
ZnO.\cite{pssbaccepted,Ronning:2014,Lorke:16,Rosa2014}. Hybrid
functionals are approximations to the exchange-correlation energy
functional in density-functional theory that incorporate a certain amount of exact exchange
from the Hartree-Fock term. The PBE0
functional\,\cite{PBE0} mixes the PBE exchange energy with the 25\% of
Hartree-Fock exchange energy treating the correlation at PBE level.
On the other hand the HSE\,\cite{HSE} exchange-correlation functional
uses an error function screened Coulomb potential to calculate the
exchange the energy. This functional mix 25\% of Hartree-Fock exchange
in the DFT exchange energy while treating the correlation part at DFT
level solely. We have used 36\%, as it is found to reproduce the
experimental gap of ZnO, as shown in
Refs.\,\cite{CRC,Janotti:09,Lany:10,pssbaccepted}.

\section*{Results and discussions}
The electronic band gaps for ZnO bulk in the wurtzite structure, bare
and hydrogenated wires are 3.6, 4.0 and 3.6\,eV, respectively.  The
value for ZnO bulk is sligtly larger than the experimental value of
3.4\,eV\,\cite{CRC}. In bare wires, surface states appear below the
bottom of conduction band due to Zn dangling bonds (around 0.5 eV) and
at the top of the valence band due to O dangling bonds, reducing the
band gap (not shown). The adsorption of hydrogenated wires supresses the surface
states and therefore produces a band gap close to the bulk
value\,\cite{XuAPL2007,Fan:07,XuPRB2009}.

\begin{figure}[htb]
\centering
\includegraphics[width=8cm,keepaspectratio]{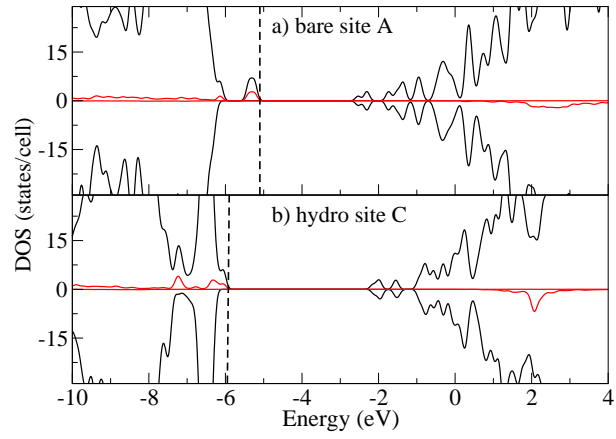}
\caption{ Density of states of  (a) Mn at surface A position in ZnO bare wires and (b) at inner C position in hydrogenated wires with HSE. The Fermi level (dashed line) is set to the highest occupied state.}
\label{fig:dos_doped}
\end{figure}

Doping of ZnO bulk with Mn splits the Mn-3$d$ states into majority $e$
and minority $t_{2}$ states, with manganese assuming a high-spin
configuration (S=5/2). Projected local magnetic moments on Mn atoms
are 5.0 $\mu_{\rm B}$.  Mn in ZnO bulk produces states around 0.5\,eV
above the valence band maximum with a strong hybridization as shown in
Fig.  \ref{fig:dos_doped} (b). 

\begin{figure}[ht!]
\centering
\includegraphics[width=8cm]{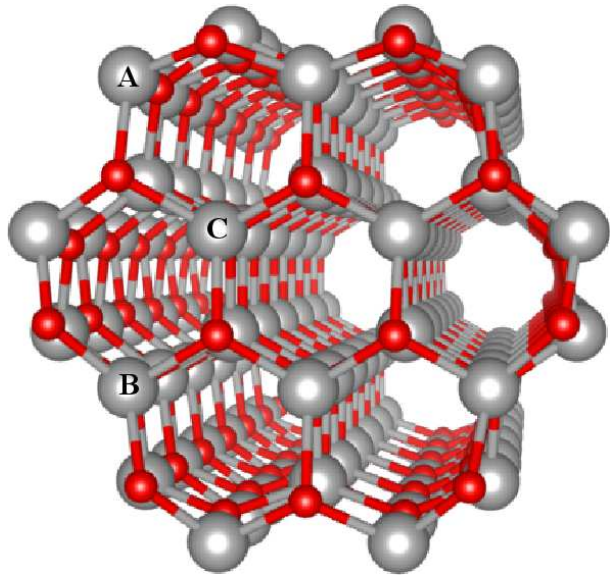}
\caption{\label{fig:geometries} Atomic configurations for ${\rm Mn_{Zn}}$ in bare ZnO nanowires. The surface A, subsurface B and inners C 
sites for Mn substitutional at Zn sites are indicated.}
\end{figure}

The wire geometries we have investigated are shown in
Figs.\,\ref{fig:geometries}. A single Mn atoms is incorporated at
surface (A), subsurface (B) and inner sites (C).  A single Mn
occupying a substitutional Zn site in the middle of the wire does not
produce strong distortion in the ZnO wire lattice. For Mn sitting at
subsurface/surface sites little relaxation is seen as well, as the covalent
radius is similar to Zn. The Mn-O bond lengths remain very close to
the values in pure ZnO, ranging from 1.8-2.1 \,{\AA} in bare wires and
from 2.0-2.2\,{\AA} in hydrogenated wires.

Incorporation of a single Mn impurity leads to site-dependent
formation energies, as shown in Table\,\ref{table:formation}.  For
bare wires, the preferred position is the surface position A. The small
volume of the nanostructure and surface effects leads to defect
migration towards the surface, as it has been discussed
\,\cite{Dalpian2013,darosa2010,Erwin2005,darosa2010,pssbaccepted,Deng2014}. On
the other hand by passivating the nanowire surfaces with hydrogen, Mn
have a lower formation energy than when it is incorporated in the
bulk position C.  The energy difference between bulk and surface position is
0.51\,eV for bare wires and 1.35\,eV for hydrogenated wires. Total
magnetic moments are not sensitive neither to site position nor to
passivation, although a small hybridization is found with nearest-neighbor
oxygen atoms.

The total and projected density-of-states (PDOS) for Mn doped wires
are shown in Figs.\,\ref{fig:dos_doped}.  We discuss only the
energetically most stable structures for both bare and hydrogenated wires. In
Fig.\,\ref{fig:dos_doped}(a) we show the results for a Mn at surface
site A in ZnO bare wire. The inpurity produces states in the ZnO band
gap located at 0.8 and 2.0\,eV above the valence band maximum
(VBM). This leads to a
decreasing in the ZnO band gap, as it has been observed in
photoluminescence experiments\,\cite{SAMADI20162,Rusdi2015}. In all
cases the minority spin states lie deep inside the ZnO CBM.

The perfect ZnO ${(10\overline 10}$ surface consists of pairs of ZnO
units. In bulk ZnO, the atoms are fourfold coordinated, but at the
surface they are three-fold coordinated exposing Zn and O dangling
bonds. When the surface is hydrogenated, as a result of reaction with
hydrogen molecules or water, the most stable situation is a full
coverage. The stability of these structures has been discussed in
Refs.\cite{Meyer2003,XuPRB2009}. In fully hydrogenated wires, Mn
states are located inside the valence band (VB) as it can be seen from
Figs.\,\ref{fig:dos_doped} (b). Additionally, majority spin states $e$ lie 1\,eV
above ZnO VBM and therefore hybridize with the O-2$p$ states.

\begin{figure}[htb]
\centering
\includegraphics*[width=8cm,keepaspectratio]{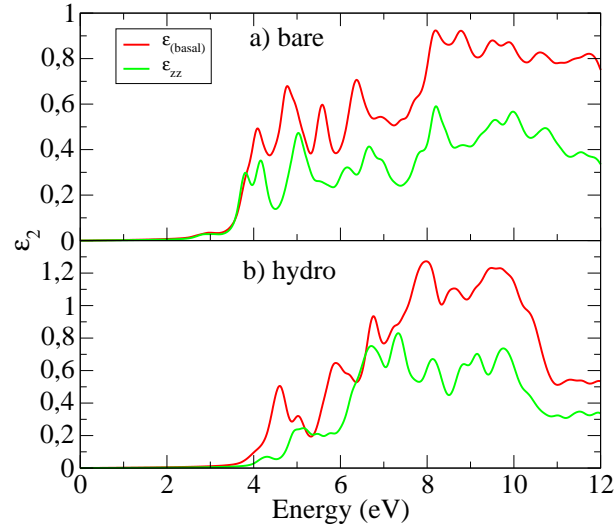}
\caption{Imaginary part of the dielectric function of Mn doped ZnO nanowires: a) Mn at surface site in bare wires and b) Mn at inner site in hydrogenated wires.
$\varepsilon_{zz}$ and (b) for $\varepsilon_{basal}$ are the field propagation along the wire growth direction and along the bas\
\
al plane.}
\label{fig:diel}
\end{figure}

Whether the optical transitions involving the intragap states are
allowed needs to be verified.  Therefore we have calculated the
dielectric function of the most stable structures, i. e., Mn at an a
surface site in bare wires and Mn at an inner site in hydrogenated
wires.  Fig. \ref{fig:diel} shows the imaginary part $\epsilon_2$ of
the dielectric function for the electromagnetic field propagation
perpendicular $\varepsilon_{\perp}$ and parallel $\varepsilon_{basal}$
to the wire growth direction.  The $\varepsilon_{\perp}$ component
shown in Fig.~\ref{fig:diel} (a) for the bare wire has an onset of 2\,eV
for absorption corresponding to the Mn state located in the band
gap shown in \ref{fig:dos_doped} (c). This means that these states are
optically active.

On the other hand $\varepsilon_{\perp}$ in hydrogenated wires shown in
Fig.\ref{fig:diel} (b) has its absorption onset located at 3.6\,eV,
which is the electronic band gap of the bare ZnO wire. This means that
these states are not optically active. In both cases the
$\varepsilon_{basal}$ component is larger. Usually ZnO nanowires grow
along the [0001] direction with the non-polar surfaces exposed as
facets. As we have recently shown in Ref.\,\cite{JPCCaccepted} that
the incidence of the external electromagnetic field along the
non-polar direction could be better for excitation response. This is
again confirmed in the present work.  In order to establish a
correlation with experimental observations, we should mention that
recently UV-vis experiments found that doping ZnO with Mn causes band
gap narrowing\,\cite{ResultsinPhysics}.  Our results show that this
corresponds to for both bare and hydrogenated wires, altough with
different values.
ZnO has been reported to have several intrinsic defects with low
formation energies, oxygen and zinc vacancies among
them\,\cite{Clark:10,Janotti:07,Erhart:05,Lany}. In nanostructures, we
have shown that O vacancies in ZnO nanowires might be present and have
low diffusion barrier towards surface sites in comparison to the
opposite path\,\cite{Deng2014}. This kind of defect can affect the
adsorption of molecules and incorporation of
dopants\cite{PCCP,Moreira_2009_AAc}. With this in mind, we have include the
presence of an oxygen vacancy in the Mn placed at inner sites of ZNO
nanowires. We have considered two situations: a vacancy placed close
to the Mn and at a vacancy at a surface site. This is motivated by
photoluminescence studies which confirmed the formation of oxygen
vacancies in Mn doped ZnO nanostructures\,\cite{Panda2016}. We find
that the presence of a vacancy in Mn doped bare wires is more stable
for a vacancy sitting at a surface site C by 1.00\,eV. This is in
agreement with our previous calculations where we predicted that
oxygen vacancies should migrate to surface sites in bare
wires\cite{Deng2014}. Interestingly enough, the magnetic moment of
these systems is not affected by the vacancy. Perhaps a more realistic
picture should be an ionized vacancy which is out of the scope of this
paper.

\begin{figure}[htb]
  \begin{tabular}{ccc}
\includegraphics*[width=0.2\textwidth,keepaspectratio]{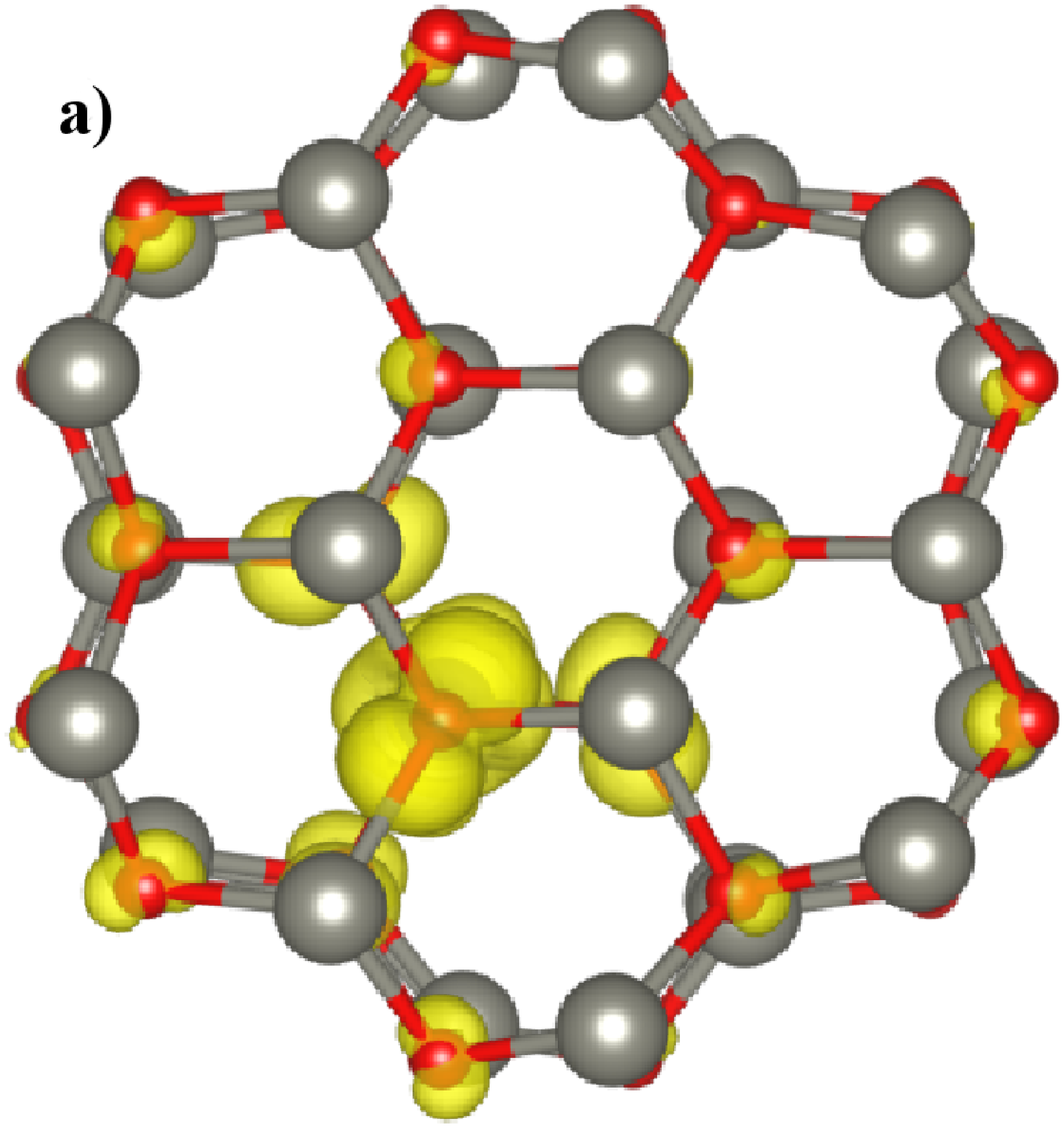} &\hfill &
\includegraphics*[width=0.2\textwidth,keepaspectratio]{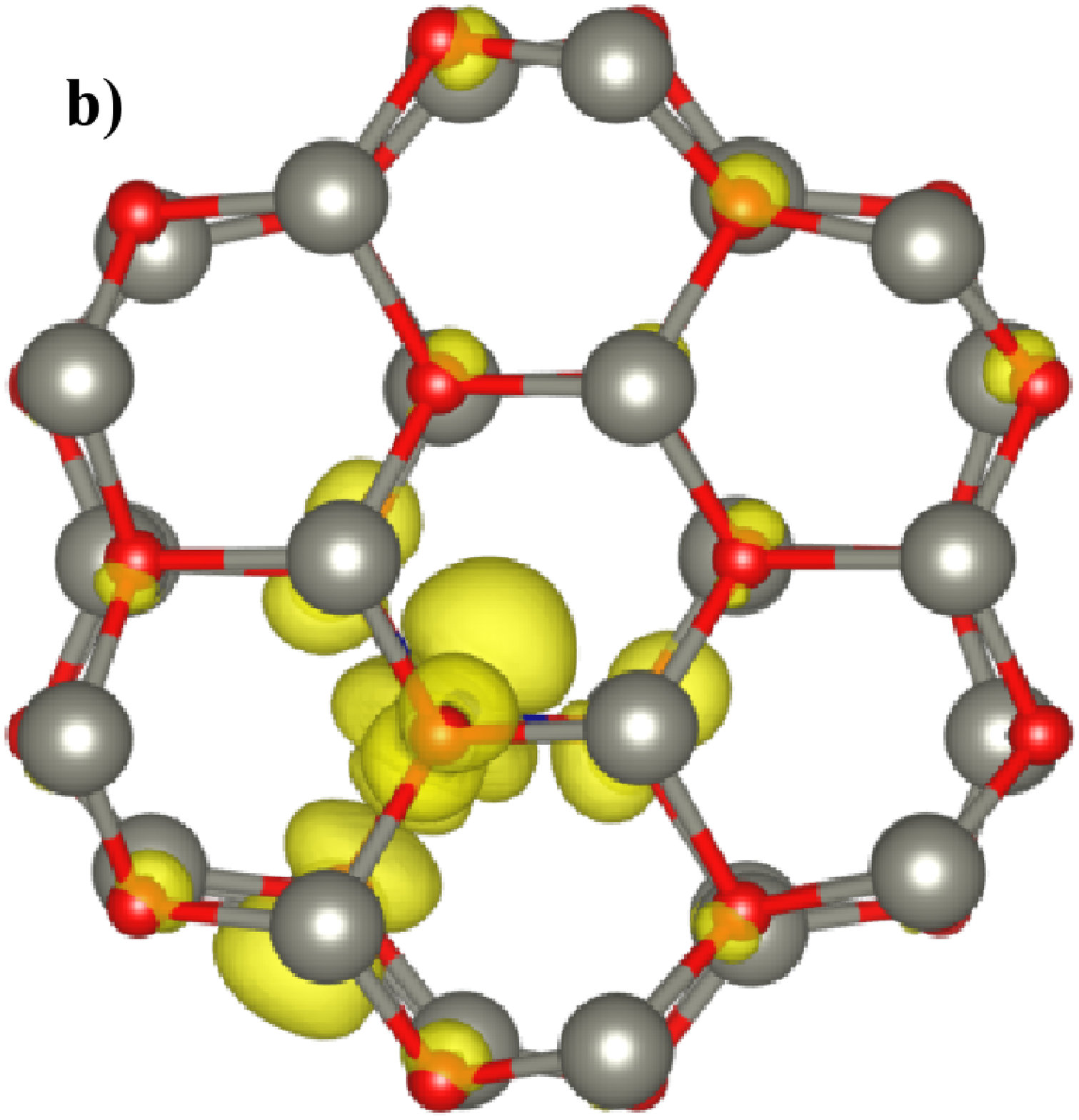}\\
\includegraphics*[width=0.21\textwidth,keepaspectratio]{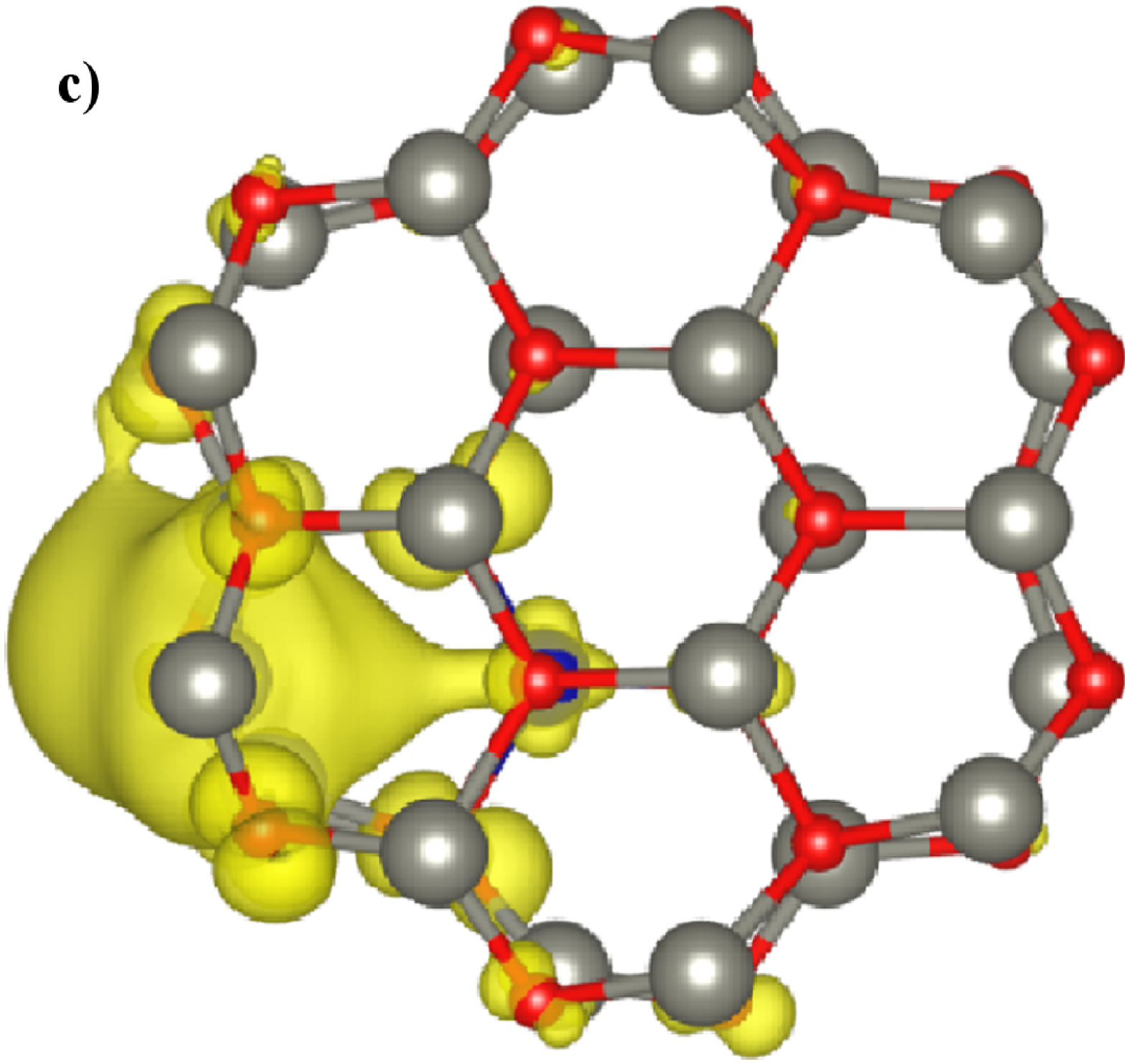} &\hfill &
\includegraphics*[width=0.2\textwidth,keepaspectratio]{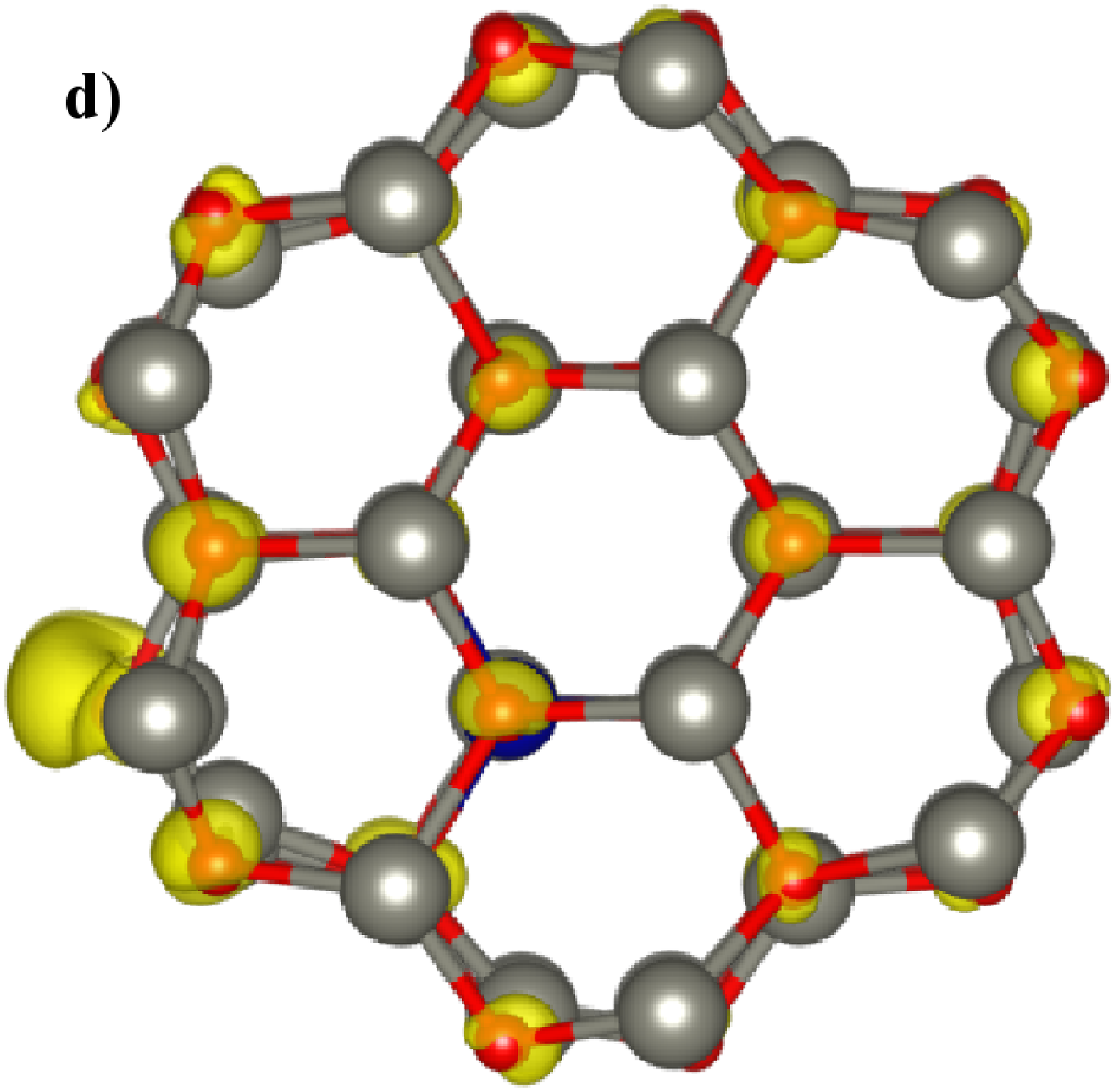}
\end{tabular}
\caption{Band decomposed charge density for Mn doped ZnO bare
  nanowires: (a) and (b) in the absence and (c) and (d) in the
  presence of an oxygen vacancy. The Mn atom is at the inner site C
  and the oxygen vacancy at the surface of the wire. The charge is
  show at the $\Gamma$ point for the highest occupied state (a) and
  (c) and for the lowest unoccupied state (b) and (d). Isosurface
  value is 0.0003 e ${\AA}^3$.}
\label{fig:parchg}
\end{figure}

In order to provide further insight on how the charge distribution in
these wires we have calculated the band decomposed charge density.
Charge localization is affected by the vacancy, but it is due to
surface dangling bonds. In Figs. \ref{fig:parchg} we show the band
decomposed charge density for Mn place at the C site of a bare ZnO
nanowires in the absence Figs. \ref{fig:parchg}(a) and (b) and in the
presence \ref{fig:parchg}(c) and (d) of an oxygen vacancy at the
surface of the wire. The charge is show at the $\Gamma$ point for the
highest occupied state (a) and (c) and for the lowest unoccupied state
(b) and (d).  In the absence of defects, the charge is localized
around the Mn atom for both the highest occupied and lowest unoccupied
states. In the presence of an oxygen vacancy, the highest occupied
state is more localized at the surface for the highest occupied
state. This charge localization has also has been found for ZnO
non-polar surfaces \cite{PCCP10,darosa2010}. The lowest unoccupied states
now has contributions from surface sites. We can conclude that there
is a stronger charge separation when a vacancy is at surface
site. This effect appears to be more important for charge separation
than the surface dangling bonds. This could be beneficial for
photocatalytic processes, since it is expected that enhanced charge
separation leads to longer lifetimes in photocatalytic
devices\,\cite{Opoku}.

\section{Conclusions}
We have investigated ZnO nanowires doped with Mn using hybrid density
functionals.  We show that the impurity prefers to sit at bulk
positions when the surface is adsorbed by hydrogen. On the other hand,
bare wires suffer from self-purification problems leading to
segregation of the dopant of Mn towards surface sites. As metal oxide
semiconductor photocatalysts are promising for fuel generation from
water splitting and carbon dioxide reduction, several strategies have
been employed to use nanostructured ZnO as photocatalyst.  In this
work we find that Mn atoms in ZnO create states in the oxide
nanostructure. However, for bare wires, these states strongly
hybridize with the surface states stemming form the surface sites and
oxygen vacancies.

\bibliography{references}
\bibliographystyle{apsrev4-1}

\section*{Acknowledgements}
We are thankful for the financial support from the Brazilian agencies
CNPq and FAPEG. A.L.R and T.F. would like to thank also German Science Foundation (DFG) under the program FO
R1616.

\begin{figure}[ht!]
\centering
\includegraphics[width=8cm]{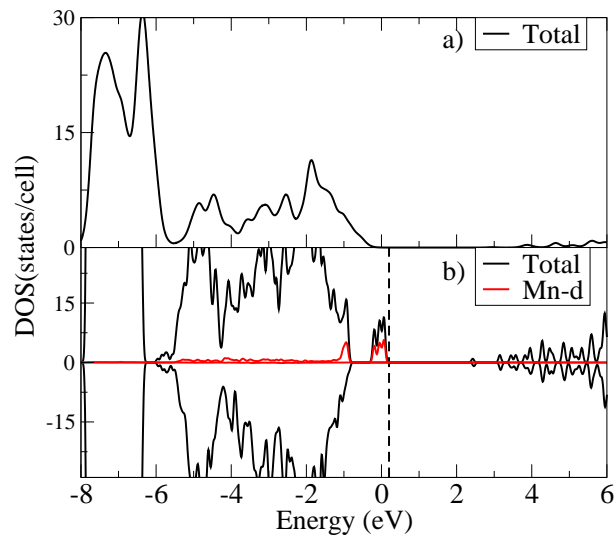}
\caption{\label{fig:dos_bulk_bare_hydro} Density of states of a) pure ZnO bulk and b) Mn doped ZnO bulk with a Mn concentration of 2.7\%. 
The dashed line represents the Fermi level.}
\end{figure}

\begin{table*}[h!]
  \small
  \caption{\label{table:formation} Total and atom projected magnetic moments $\mu_{\rm tot}$ (in $\mu_{\rm B
})$. Relative formation  energies $\rm E_f$ (in eV) of
  neutral Mn impurities in ZnO calculated with HSE.}
\begin{tabular*}{\textwidth}{@{\extracolsep{\fill}}lllllllll}
\hline
site      & \multicolumn{4}{c}{bare} & \multicolumn{4}{c}{hydrogenated}\\
\hline
       &   $\mu_{\rm tot}$  &  $\mu_{\rm Mn}$  &  $\sum{\mu_{\rm O}}$ & $\rm E_f$ & $\mu_{\rm tot}$  &   $\mu_{\rm Mn}$   &  $\sum{\mu_{\rm O}}$ & $\rm E_f$ \\
\hline
inner  &    5.0       & 4.3 &   0.09  & 1.35   & 5.0 & 4.5  & 0.12 & 0.00\\
sub     &   5.0      & 4.4  &  0.12   & 0.13   & 5.0   &  4.5 & 0.11 & 0.50\\
surf    &   5.0       & 4.4   & 0.10    & 0.00   & 5.0  & 4.5 & 0.07 & 0.51 \\
\hline         
\end{tabular*}
\end{table*}

\end{document}